\documentclass[12pt]{article}
\usepackage{geometry}
\usepackage{amsmath}
\usepackage{amssymb}
\usepackage[round]{natbib}
\usepackage{graphicx}
\usepackage[T1]{fontenc}
\usepackage[utf8]{inputenc}
\usepackage{authblk}
\usepackage{setspace}
\title{Step selection techniques uncover the environmental predictors of space use patterns in flocks of Amazonian birds}
\author[a]{Jonathan R. Potts\thanks{jrpotts@ualberta.ca}}
\author[b,c]{Karl Mokross}
\author[b,c]{Philip C Stouffer}
\author[a,d]{Mark A. Lewis}
\affil[a]{Centre for Mathematical Biology, Department of Mathematical and Statistical Sciences, University of Alberta, Canada}
\affil[b]{School of Renewable Natural Resources, Louisiana State University Agricultural Center, Baton Rouge, Louisiana, 70803}
\affil[c]{Projeto Din\^amica Biol\'ogica de Fragmentos Florestais. INPA. Av. Andr\'e Ara\'ujo 2936. Petr\'opolis. Manaus. Brazil. 69083-000}
\affil[d]{Department of Biological Sciences, University of Alberta, Edmonton, Canada}

\date{*Corresponding author: jrpotts@ualberta.ca}
\begin{document}
\maketitle
\section*{Summary}
\begin{enumerate}
% The background to the question (why it is interesting)
  \item Anthropogenic actions cause rapid ecological changes, meaning that animals have to respond before they have time to adapt.  Tools to quantify emergent spatial patterns from animal-habitat interaction mechanisms are vital for predicting the population-level effects of such changes.   
% What the question is
\item  Environmental perturbations are particularly prevalent in the Amazon rainforest, and have a profound effect on fragmentation-sensitive insectivorous bird flocks.  Therefore it is important to be able to predict the effects of such changes on the flocks' space-use patterns.
%  What was done in the study
\item We use a step selection function (SSF) approach to uncover environmental drivers behind movement choices.  This is used to construct a mechanistic model, from which we derive predicted utilization distributions (home ranges) of flocks.
%  What was found
\item We show that movement decisions are significantly influenced by canopy height and topography, but not resource depletion and renewal.  We quantify the magnitude of these effects and demonstrate that they are helpful for understanding various heterogeneous aspects of space use.  We compare our results to recent analytic derivations of space use, demonstrating that they are only accurate when assuming that there is no persistence in the animals' movement.
%    What this means in the context of the broad field of animal ecology
\item Our model can be translated into other environments or hypothetical scenarios, such as those given by proposed future anthropogenic actions, to make predictions of spatial patterns in bird flocks.  Furthermore, our approach is quite general, so could be used to predict the effects of habitat changes on spatial patterns for a wide variety of animal communities.
%   \ldots
   \end{enumerate}
\section*{Keywords}
%Listed in alphabetical order, the key-words should not exceed 10 words or short phrases. Please pay attention to the keywords you select: they should not already appear in the title or abstract. Rather, they should be selected to draw in readers from wider areas that might not otherwise pick up your paper when they are using search engines.
Amazon rainforest; Animal movement; Deforestation; Forest fragmentation; Home range; Insectivorous birds; Resource selection; Space use; Step selection; Theoretical ecology
\section*{Introduction}
%\linenumbers
%This should state the reason for doing the work, the nature of the hypothesis or hypotheses under consideration, and should outline the essential background. Of course, cite many relevant works here and throughout \citep{harrisonetal10,Hubbellbook,Gergsthesis}.

Today's rapid and extensive changes to the environment highlight the need for accurate models that can predict the effect of these perturbations on animal and plant populations \citep{thomasetal2004}.  The Amazon rainforest is a prime example of such a disturbed ecosystem, with recent large-scale deforestation causing a wide range of ecological changes \citep{lauranceetal2004, fearnside2005}.  In particular, insectivore bird communities living in the rainforest find it difficult or impossible to survive in deforested areas \citep{sekercioglusetal2002}, even after years of second growth regeneration \citep{sodhietal2008}.  This fragmentation sensitivity ultimately forces many bird species into local extinction \citep{stoufferetal2011}, reducing the biodiversity and consequently ecosystem functioning \citep{chapinetal2000,mantylaetal2011}.

Mixed-species flocks are an important element of the understory avifauna found in practically all {\it terra firme} forests in the Amazon basin.  The flocks tend to comprise of around 20 to 60 insectivore species that actively forage in the vegetation \citep{munn1984,powell1989,mokrossetal2014}, searching the different strata and substrates in the vegetation for practically the whole duration of the day with high consumption rates (K. Mokross pers. obs.).  This makes them not only important contributors to the species richness of the Neotropical avifauna \citep{powell1989}, but also potential key players in trophic cascades with herbivorous insects and plants \citep{mantylaetal2011}. 

The core of each flock is composed of 5-10 species which are consistently present, and share the same overlapping territory \citep{munnterborgh1979}, each pair defending its territory from conspecifics.  In the flocks studied here, the Cinerous Antshrike ({\it Thamnomanes caesius}) plays a nuclear role by giving alarm and rally calls that maintain flock cohesiveness \citep{munn1986}.  Other species will be frequent flock attendants but leave occasionally, either by switching between flocks, or by having smaller territories than the core species \citep{jullienthiollay1998}.

Space use for these flocks is very stable: territories' shapes change little in two decades \citep{martinezgomez2013}.  The core species gather in the same location at dawn every day, usually in a central position within its territory, and will begin foraging from there in an apparently random fashion \citep{powell1985} until sunset where they roost in the approximate vicinity to each other  \citep{jullienthiollay1998, martinezgomez2013}. The purpose of this paper is to disentangle the purposeful drivers behind the movement processes from this apparently random behavior, then to use this understanding to build a predictive model of space use patterns in insectivore bird flocks.

% Importance of relating movement to space use in order to predict the effects of changes
Linking animal movement to space use in a quantitative, analytic fashion is vital for predicting the effects of environmental changes on animal populations \citep{moralesetal2010}.  The factors driving the animals' movement ultimately determine the size and structure of the space that they use in order to meet their everyday needs.  By uncovering how these movement processes give rise to spatial patterns, it would be possible to predict the types and amounts of terrain that would be used were the environment to be perturbed, by anthropogenic effects or otherwise  \citep{nathanetal2008}.

% Three hypotheses we test
In this paper, we make an important step towards this end, by identifying and quantifying some of the key environmental factors that influence Amazonian bird flock movement, then using them to construct a predictive model of space use.  Our approach begins by using a step-selection function \citep{fortinetal2005} to test three hypotheses regarding the drivers behind the flocks' movement decisions.  Then we derive from this a master equation to link these processes to the emergent space use patterns.  The hypotheses we test are that (1) flocks are more likely to move into areas with taller canopies than shorter, (2) flocks tend to move away from higher ground and towards lower, (3) flocks leave some time for the resources to renew before re-visiting a tree they have recently visited.

Taller canopies are expected to be preferable for birds since they tend to contain a larger mass of resources \citep{bassetetal1992}.  On the other hand, lower ground can be support more buffered conditions from wind turbulence and sunlight from outside the forest cover \citep{ewersbanksleite2013}, and naturally hold higher air and soil moisture levels \citep{baralotocouteron2010} which could potentially increase arthropod loads per vegetation volume \citep{williamslineraherrera2003, chanetal2008}.  

We begin with these two in order to develop a basic methodological framework that we can easily extend to build more complicated models that could include other factors driving the birds' movement decisions, such as memory \citep{smouseetal2010}, territoriality \citep{PHG3}, or tighter movement patterns in dense foliage \citep{jullienthiollay1998}.  Proceeding in this way is advantageous since we gain a clear understanding of exactly how, and to what extent, each environmental factor influences flock movement.  Though starting with a more complex model may lead to more accurate predictions, it would make it harder to disentangle the relative effects of each model parameter on the resulting space use, so we would ultimately learn less about the underlying causes of space-use phenomena.

% State that we compare this to Moorcroft-Barnett predictions
Recently, \citet{moorcroftbarnett2008} derived an approximate analytic formula relating an animal's utilization distributions to its step selection mechanism, in a particular 1-dimensional scenario.  This was then extended into arbitrary dimensions by \citet{barnettmoorcroft2008}.  We examine to what extent this prediction is accurate for our more complicated scenario, by examining where the analysis breaks down as models become more complex.  We compare their predictions with numerical solutions of our model to demonstrate where their results hold to a good degree of accuracy, so can be usefully applied to empirical data, and where they fail.

\section*{Materials and methods}
%This should provide sufficient details of the techniques to enable the work to be repeated. Do not describe or refer to commonplace statistical tests in Methods but allude to them briefly in Results.
\subsection*{The step selection function model}

% How we tested the three hypotheses: construction of a step-selection function using weights, turning angles and step lengths
Our model for bird-flock movement is based on a step selection function (SSF) approach \citep{fortinetal2005}.  Following the formalism initiated by \citet{rhodesetal2005}, 
%then extended by \citet{PBMSL} 
but extended here to take into account correlations in the movement, we write the probability $f({\bf x}|{\bf y},\theta_0)$ of finding an animal at position ${\bf x}$, having traveled from ${\bf y}$ in the previous step, given that it arrived at ${\bf y}$ on a bearing of $\theta_0$, as follows 
\begin{equation}
f({\bf x}|{\bf y},\theta_0) = \frac{\Phi({\bf x}|{\bf y},\theta_0)w({\bf x},{\mathcal E})}{\int_\Omega {\rm d}{\bf x}' \Phi({\bf x}'|{\bf y},\theta_0)w({\bf x}',{\mathcal E})}.
\label{ssf}
\end{equation}
Here, $w({\bf x},{\mathcal E})$ is a weighting function that depends upon the animal's position ${\bf x}$ and some environmental covariates ${\mathcal E}$ \citep{foresteretal2009}, $\Phi({\bf x}|{\bf y},\theta_0)$ is the probability of being at ${\bf x}$ in the absence of habitat selection, given that the animal was previously at ${\bf y}$ and had arrived there on a bearing of $\theta_0$, $\Omega$ is the study area, and bearings are measured in an anti-clockwise direction from the right-hand half of the horizontal axis.  Each step takes a fixed amount of time $\tau$.  The function $\Phi({\bf x}|{\bf y},\theta_0)$ allows us to take into account the fact that animals may be more likely to take steps of a particular length, and the distribution of such lengths can be derived from empirical data.  For computational purposes, we truncated the step length distribution at steps of greater than 100$m$, since these never occur in our data.  We include the angle $\theta_0$ into this formulation to allow for the possibility of correlations between successive movement bearings.

For the purpose of testing hypotheses (1) and (2), $w({\bf x},{\mathcal E})$ is a function of the canopy height $C({\bf x})$ and the topography (i.e. elevation above sea level) $T({\bf x})$, both measured in meters $(m)$.  We test two candidate formulations for $w({\bf x},{\mathcal E})$ 
\begin{equation}
w_a({\bf x},\alpha,\beta)=\exp[\alpha C({\bf x})-\beta T({\bf x})],
\label{weight_fn_a}
\end{equation}
\begin{equation}
w_b({\bf x},\alpha,\beta)= C({\bf x})^\alpha T({\bf x})^{-\beta}.
\label{weight_fn_b}
\end{equation}
Notice that Eq. \ref{weight_fn_b} can also be written as $w_p({\bf x})=\exp\{\alpha \ln[C({\bf x})]-\beta \ln[T({\bf x})]\}$, in keeping with the original formulation of the step selection function from \citet{fortinetal2005}.  Since we would expect the birds to be more likely to move towards lower ground than higher, we place a minus sign before the $\beta$ in each equation, so that $\beta$ is expected to be positive.  

To test hypothesis (3), we assume that the resource amount at the start of the day ($t=0$), is proportional to the canopy height \citep{camposetal2006}.  As the birds move through an area, they deplete the resources, which take a time $G\tau$ to renew.  The resource amount present at a site at time $g\tau$ after having been visited is assumed to be $R({\bf x},t,G)=gC({\bf x})/G$ as long as $g<G$, otherwise $R({\bf x},t,G)=C({\bf x})$.  Here, $t$ is the time since start-of-day and a unit of resources is implicitly defined to be the maximum amount of usable resources sustainable by a tree per meter of tree height.  At time $t=0$, we assume $R({\bf x},0,G)=C({\bf x})$.  As with hypotheses (1) and (2), we test two candidate formulations for $w({\bf x},{\mathcal E})$  
\begin{equation}
w_c({\bf x},t,\alpha,\beta,G)=\exp[\alpha R({\bf x},G,t)-\beta T({\bf x})],
\label{weight_fn_c}
\end{equation}
\begin{equation}
w_d({\bf x},t,\alpha,\beta,G)= R({\bf x},t,G)^\alpha T({\bf x})^{-\beta}.  
\label{weight_fn_d}
\end{equation}
Notice that when $G=1$, we have $w_a({\bf x},\alpha,\beta)=w_c({\bf x},t,\alpha,\beta,1)$ and $w_b({\bf x},\alpha,\beta)=w_d({\bf x},t,\alpha,\beta,1)$.

\subsection*{Data collection methods}

% A section on how the data was collected
Flock activity is conspicuous, allowing birds to be followed on foot.  As flocks moved, geolocations were recorded at 30 second intervals with a hand-held GPS unit (Garmin Vista HCX). The observer maintained a distance of 10-20m from the flocks to ensure no alarm or avoidance behavior was induced in the birds.  Observer distance is not in perfect lockstep with the flock, yet the average distance to the approximate center of the flock could be maintained to an accuracy of a few meters.  Where possible, we used the location of a Cinereous Antshrike as the flock location.  This species was usually conspicuous in the center of the flock.  Other more active species typically spread out over a radius of 5-10 m, depending on the size and speed of the flock.  

Though GPS error can be around $10m$, it is mainly caused by the relatively slow movement of the ionosphere \citep{parkinsonspilker1996} which only changes by a few centimeters during 30 second intervals.  Indeed, evidence from using hand-held GPS for tracking butterflies suggests that the median drift (i.e. absolute error) between consecutive 15 second relocations is only $8cm$ \citep{severnsbreed2014}.  Therefore it is reasonable to assume that the measured step lengths and turning angles accurately reflect reality.

Compared to other available methods, this data reflects well the movement of flocks on a small spatio-temporal scale. It provides a high resolution of time sequence that is not possible in radio-telemetry studies, and presently no other techniques allow the gathering of detailed spatial data for passerines of this size.  Unlike remote telemetry, this method also allows the direct observation of behavior, so the observer can directly verify whether the recorded spatial locations are corresponding well with the canopy height and topographical maps.

For measuring canopy heights, we used a LIDAR (Light Detection and Ranging) canopy height model (CHM).  Similarly, topography (Digital Elevation Models DEM) was acquired using small footprint airborne LIDAR.  The derived (post-processed) images from the LIDAR data are 1m/pixel resolution, which we transformed into 10m lattices by bilinear interpolation.  LIDAR data were collected by airborne laser scanning using a Hexagon-LEICA ALS50 PHASE II  MPiA sensor of 150 kHz, at 800 m altitude, with 24 degrees opening, 118 MHz pulse rate, 58 Hz scan rate, 3,7 points/$m^2$ density. Swaths were of 340 meters wide, spaced at 240 meters. Post-processing used a Forest Service methodology to generate DEM and CHM at 1 $m^2$/pixel [see \citet{starketal2012} for more details on LIDAR data collection and analysis].

We analyzed movement of 6 different flocks in {\it terra firme} forest, during the dry seasons between June and November during 2009-2011, at the Biological Dynamics of Forest Fragments Project (BDFFP), about 70 km north of Manaus, Brazil (see http://pdbff.inpa.gov.br/ for maps).  The study area is a rectangle of size $2.8km$ by $1.5km$.  Flocks were each tracked for between 5 and 11 days.  Each flock gathers in one particular place each day, then moves around the forest for a total of about eleven-and-a-half hours during the day, before each bird goes back to its roost for the night.  %The flock positions were recorded every 30 seconds during the time that they were active.  

%The specs for the LIDAR raw data acquisition (just in case):

%-Flight altitude: 800 m
%-Scan opening angle: 24 degrees
%-Pulse rate: 118 kHz;
%- Scan rate: 58 Hz;
%- Point density (DEM): 3,7 pts/m�;
%- DEM (all points) with an irregular  grid  of points spaced approximately 
%0,52 meters (average)
%-  XY precision: 0,30 meters to 1 sigma (68%)
%- Z precision : 0,20 meters to 1 sigma (68%)

\subsection*{Parametrizing the models from the data}

% Step length and turning angle distributions
% Using 1 min and 5 min data
The first step in parametrizing the models is to calculate the step length and turning angle distributions, i.e. the distance between successive positions and the angle an animal turns through from one move to another, respectively [see e.g. \citet{cristetal1992}].  Since these depend upon the temporal resolution $\tau$ (i.e. the time between successive position fixes), we use both $\tau=1\mbox{ minute}$ and $\tau=5\mbox{ minutes}$, deriving two different sets of step length and turning angle distribution for the different values of $\tau$.  The value $\tau=1\mbox{ minute}$ is chosen because bird flocks tend to move from one tree to another at an average of approximately every 1 or 2 minutes.  Though their movement is a continuous rather than discrete process, the model is formulated so this timescale roughly represents the small-scale decisions that the birds make regarding whether they stay in a tree or choose to move to another.  We also examine the case $\tau=5\mbox{ minutes}$ to determine whether the decisions about where to move can instead be viewed as taking place on a timescale longer than a single jump between trees.  In other words, the birds might only be considering the next tree they move to when deciding where to go ($\tau=1\mbox{ minute}$), or they might be thinking a few trees ahead when they make this decision ($\tau=5\mbox{ minutes}$).

The step length distributions are fitted to both a Weibull distribution \citep{foresteretal2009} and an Exponentiated Weibull (EW) distribution \citep{nassareissa2003}, using the Akaike Information Criterion (AIC) to determine the best model, whereas we fit the turning angles to a von-Mises distribution \citep{marshjones1988}.  The Weibull, EW and von Mises distributions have the following forms, respectively

\begin{equation}
\rho_1(x|a,b) = \frac{a}{b}\left(\frac{x}{b}\right)^{a-1}\exp\left[-\left(\frac{x}{b}\right)^a\right],
\label{step_length_weibull}
\end{equation}
\begin{equation}
\rho_2(x|a,b,c) = \frac{ac}{b}\left(\frac{x}{b}\right)^{a-1}\exp\left[-\left(\frac{x}{b}\right)^a\right]\left\{1-\exp\left[-\left(\frac{x}{b}\right)^a\right]\right\}^{c-1},
\label{step_length_exp_weibull}
\end{equation}
\begin{equation}
V(\phi|k) = \frac{\exp[k\cos(\phi)]}{2\pi\mbox{I}_0(k)}.
\label{turning_angles}
\end{equation}

% The testing grid
Since the rainforest canopy consists of distinct treetops whose widths are each roughly $10m$ across, we split the terrain $\Omega$ into a grid $S$ of $10m$ by $10m$ squares.  This allows us to associate a value of $C(s)$ and $T(s)$ to each square $s$ in $S$, respectively the mean canopy height and mean topography of the square.  Parametrizing Eq. \ref{ssf} from the data therefore requires maximizing the following likelihood function

\begin{equation}
L({\bf X}|{\mathcal E}) = \prod_{n=2}^N \frac{\Phi({\bf x}_n|{\bf x}_{n-1},\theta_{n-1})w({\bf x}_n,{\mathcal E})}{\sum_{s \in S} \Phi(s|{\bf x}_{n-1},\theta_{n-1})w(s,{\mathcal E})}, 
\label{likelihood_fn}
\end{equation}

where ${\bf X}=\{{\bf x}_0,{\bf x}_1,\dots,{\bf x}_N\}$ are the consecutive positions of a flock, $\theta_{n}$ is the bearing from ${\bf x}_{n-1}$ to ${\bf x}_n$, $\Phi$ is the product of the best-fit step length and turning angle distributions, and $w$ is either $w_a$, $w_b$, $w_c$ or $w_d$, depending on which model we are fitting.  

To test hypothesis (1), we fix $\beta=0$ and find the value of $\alpha$ that maximizes $L({\bf X}|{\mathcal E})$, which we call $\alpha_{\mbox{m}}$.  We then use the likelihood ratio test to compare the resulting value of $L({\bf X}|{\mathcal E})$ with the value of $L({\bf X}|{\mathcal E})$ when both $\alpha$ and $\beta$ set to zero.  For hypothesis (2), we fix $\alpha=\alpha_{\mbox{m}}$ and find the value of $\beta$ that maximizes $L({\bf X}|{\mathcal E})$, again using the likelihood ratio test to compare this value of $L({\bf X}|{\mathcal E})$ with the one where $\alpha=\alpha_{\mbox{m}}$ and $\beta=0$.  We then find the values of $\alpha$ and $\beta$ that maximize $L({\bf X}|{\mathcal E})$ by varying both parameters simultaneously, giving best fit values denoted by $\alpha_{\mbox{bf}}$ and $\beta_{\mbox{bf}}$.  We use a Markov bootstrap method with 100 bootstraps to find standard errors for $\alpha$ and $\beta$ \citep{horowitz2003}.

Hypothesis (3) is tested by fixing $\alpha=\alpha_{\mbox{bf}}$ and $\beta=\beta_{\mbox{bf}}$ and finding the value of $G$ that maximizes $L({\bf X}|{\mathcal E})$, then using the likelihood ratio test to compare the resulting value of $L({\bf X}|{\mathcal E})$ with the value of $L({\bf X}|{\mathcal E})$ when $G=1$.  For each maximization calculation, we use the Nelder-Mead simplex algorithm \citep{Nelder_Mead}, as implemented in the Python \texttt{maximize()} function from the SciPy library \citep{scipy}.  
%Each run of the algorithm takes about 12-24 processor hours on a desktop computer, depending on the speed at which the algorithm converges.

\subsection*{Constructing the space use distribution}

% Some details of simulations: include gathering points and methods for calculating KDEs
We use two methods for constructing the space use distribution from the parametrized SSF (Eq. \ref{ssf}), via simulation analysis and through constructing the master equation and numerically deriving its steady-state solution.  For the former approach, we simulate one particular flock's movement on the grid $S$ using the jump probabilities given by SSF.  Since the flock gathers in one particular place each day, and moves around the terrain for a total of about eleven-and-a-half hours during the day, we start the simulated birds at the gathering point and run the simulation for 138 time steps, each step representing $\tau=5\mbox{ minutes}$ (giving 11 hours 30 minutes in total), taking a note of all the positions at which the flock landed after each step.  We repeat this 100 times, representing 100 days, giving 13,800 simulated positions in total.  In the data, we tend to have around 10 days per flock.  However, we use 100 here to average out some of the stochasticity.  From these simulated positions, we calculate the 50\%, 60\%, 70\%, 80\%, and 90\% Kernel Density Estimators (KDEs), using a fixed kernel method with smoothing parameter $h=\sigma n^{-1/6}$ where $\sigma = (1/2)\sqrt{\sigma_x^2+\sigma_y^2}$ and $\sigma_x, \sigma_y$ are the standard deviations of the simulated data in the $x$- and $y$-directions respectively \citep{worton1989}. % Simulations are coded in C and take about 15 seconds to run for all 100 days on a desktop computer.  
KDE calculations are performed using Python.% and take 14 minutes.

% Master equation construction
In addition to simulation analysis, we also construct the master equation for the probability density function $u({\bf x},\theta,t)$ of the animal being at ${\bf x}$ at time $t$ having traveled there on a bearing of $\theta$.  This allows us to compare our results with the predictions of \citet{barnettmoorcroft2008}, who mathematically analyzed the step selection function (Eq. \ref{ssf}) in the simpler case where the turning angle distribution is uniform.  They proved that $u({\mathbf x})$ is proportional to $w({\mathbf x},{\mathcal E})z({\mathbf x},{\mathcal E})$, where $z({\bf x},{\mathcal E})=\int_\Omega {\rm d}{\bf x}' \Phi({\bf x}'|{\bf x},\theta_0)w({\bf x}',{\mathcal E})$ is a local averaging of $w({\mathbf x},{\mathcal E})$.  We examine to what extent this result extends to our more complicated situation of a correlated random walker.  
%Following the methods of \citet{PBMSL}, we 
We use Eq. \ref{ssf} to construct the following master equation 
\begin{equation}
u({\bf x},\theta,t+\tau) = \int_{-\pi}^{\pi}{\rm d}\theta_0 \int_{0}^{r_{\rm max}}{\rm d}r \frac{\Phi({\bf x}|y_{\theta}(r),\theta_0)w({\bf x},{\mathcal E})}{\int_\Omega {\rm d}{\bf x}' \Phi({\bf x}'|y_{\theta}(r),\theta_0)w({\bf x},{\mathcal E})}u(y_{\theta}(r),\theta_0,t),
\label{ME}
\end{equation}
where $y_\theta(r)$ describes the locus of points ${\bf y}$ upon which the animal could approach ${\bf x}=(x_1,x_2)$ at bearing $\theta$, i.e. $y_\theta(r)=(x_1+\cos(\theta+\pi)r, x_2+\sin(\theta+\pi)r)$, with $r$ denoting the distance between $y_\theta(r)$ and ${\bf x}$ \citep{PBMSL}.  Here $r_{\rm max}$ is the distance along this line from $x$ to the boundary of $\Omega$ and so gives the upper endpoint of integration.  To calculate the steady-state distribution, we solve Eq. \ref{ME} iteratively until $|u({\bf x},\theta,t+\tau)-u({\bf x},\theta,t)|<10^{-8}$ for every value of ${\bf x}$ and $\theta$.  The area $\Omega$ for this calculation is defined to be the 95\% KDE of the flock used for the simulations.  We used zero-flux boundary conditions, which models the fact that the birds are confined within their territory.  Calculations were coded in C and it took approximately 2 hours to find a single steady state distribution.

Note that in these methods, we are separating the fitting of the turning angle and step length distributions from the fitting of the weighting functions.  This makes the maximization procedure far faster and means the algorithms are more likely to converge to the global maximum.  However, if the weighting function $w$ gives a particularly strong selection for an environmental covariate and/or the step length distributions are fat-tailed, then this separation may cause inaccuracies in the resulting model.  To test that this is not the case, we calculated the mean and standard deviation of the step length and turning angle distributions from the above simulations to verify that the weighting function had not significantly altered them.

\section*{Results}
%This should state the results, drawing attention in the text to important details shown in tables and figures.
\subsection*{Step length and turning angle distributions}

% Results of the step length and turning angle distributions determined from the data
For both cases $\tau=1\mbox{ minute}$ and $\tau=5\mbox{ minutes}$, the best fit step length distribution is an Exponentiated Weibull (EW) distribution (Fig. \ref{slta}).  For $\tau=1\mbox{ minute}$, $\Delta AIC = 126.9$ between EW and Weibull.  For $\tau=5\mbox{ minutes}$, $\Delta AIC = 14.6$. %the AIC for the EW distribution is $57845.3$ as compared with $57972.2$ for the Weibull distribution ($\Delta AIC = 126.9$).  For $\tau=5\mbox{ minutes}$, the AIC for EW is $14018.0$, compared with $14032.6$ for the Weibull distribution ($\Delta AIC = 14.6$).  Both these improvements in AIC give very strong evidence that the EW is a better fit than the Weibull distribution.  

The step length distributions both increase from $0m$ initially, before decaying (Fig. \ref{slta}).  However, this is not an indicator that birds are more likely to move a medium length distance than a very short distance, but is simply due to there being less area in the annulus of radius between $r$ and $r+\delta r$ when $r$ is smaller.  If $\delta r$ is small then the total amount of area into which a flock can move, given that it moves a distance between $r$ and $r+\delta r$, is approximately $\delta r \times 2 \pi r$, which is proportional to $r$.  To find the relative preferences of the birds to move a particular distance, it is therefore necessary to divide the probability density, $P(r)$, by the distance moved, $r$.  If we do this for our data on the 1 minute temporal resolution, we find that $P(r)/r$ is approximately $0.044\exp(-r/4.75)$ and for the 5 minute time-scale $P(r)/r \approx 0.0080\exp(-r/11.3)$, both of which decay monotonically as $r$ increases.

\subsection*{Hypothesis testing}

% Results of the hypothesis testing, including the MLE parameter estimates used
The tests indicate that there is a significant effect of both canopy height (hypothesis 1) and topography (hypothesis 2) on the flocks' movement.  However, accounting for resource renewal, so that birds are less likely to re-visit trees that they have recently visited, does not improve the model fit (hypothesis 3).  The conclusions are the same both for $\tau=1\mbox{ minute}$ and $\tau=5\mbox{ minutes}$, so we cannot conclude anything about the temporal resolution on which decisions are made.  Table \ref{hypothesis_test_results} gives a summary of the results.  

To put these in a biological context, consider two trees, equally accessible over a 5 minute interval and on ground of equal elevation, but one $A$\% taller than the other, e.g. if one is 30m high and the other 20m high then $A=50$.  Then the birds are $(1+A/100)^{0.277}=1.5^{0.277}\approx 1.096$ times more likely to move to the taller tree than the shorter, i.e. about 10\% more likely.  Conversely, suppose that both trees are of equal height but one tree is ground $B$\% higher above sea-level than the other.  Then the birds are $(1+B/100)^{1.697}$ times more likely to move to the tree on lower ground.  For example, an decrease from 50m to 40m elevation leads to a $1.25^{1.697} \approx 1.460$ increase in probability of moving there, i.e. they are 46\% more likely to move to the 40m elevation.

The weighting function $w_b$ (Eq. \ref{weight_fn_b}) provides a better fit to the data than $w_a$ (Eq. \ref{weight_fn_a}) for $\tau=5\mbox{ minutes}$.  The AIC for $w_b$ is %6926.9 
lower than that %6930.7 
for $w_a$ ($\Delta AIC = 3.8$).  Though the AIC for $w_b$ %(AIC = 24848.0) 
for $\tau=1\mbox{ minutes}$ is slightly lower than for $w_a$ ($\Delta AIC = 0.1$), %(AIC=24848.1), 
the change in AIC is not large enough to be considered good evidence that $w_b$ is better than $w_a$.  In Table \ref{hypothesis_test_results} we detail the results for the function $w_b$ and its generalization $w_d$ (Eq. \ref{weight_fn_d}).  Results for $w_a$ and $w_c$ (Eq. \ref{weight_fn_c}) are qualitatively similar.

\subsection*{Space use distributions}

% Plots of simulations and brief explanation: why they are a bit too big
Figure \ref{sim_plots} compares the simulated space use with the empirical data on flock positions.  The KDE contour lines for the simulated data are quite tightly packed around the edge of the empirical data points, suggesting that the model is giving a reasonable prediction of space use patterns.  However, the extent of the simulated home range is clearly larger than the empirical home range.  %This is likely due to the fact that these bird flocks are territorial, so will have pressure from neighboring flocks, confining their home range.  There may also be memory effects confining the space use \citep{briscoeetal2002}.  To improve the space use predictions would likely require extending our model to include either or both of these effects.

Though separating the fitting of the step length and turning angle distributions from the environmental interactions may mean that the fit is less accurate than if all parameters were fitted together, it turns out that the mean of the simulated data's step length distribution is $20.05\pm 0.95m$ (95\% confidence intervals), compared with $20.09m$ from the data.  The standard deviation of the simulated step lengths is $13.55\pm 2.01m$ as compared with $13.23m$ from the data.  Similarly, the standard deviation of the turning angles from simulation output is $82.1\pm8.7$ degrees as compared with $82.7$ degrees from the data, and the mean is $-0.2\pm 6.9$ degrees, as compared with $-1.7$ degrees from the data.  Therefore including the weighting function does not significantly change the step length or turning angle distributions.

\subsection*{Comparison with analytic results}

% Comparison with Moorcroft and Barnett results: where and why it breaks down
Previous work showed that if there is no correlation in an animal's movement, the steady-state space-use distribution is proportional to $w({\bf x},{\mathcal E})z({\bf x},{\mathcal E})$ as long as the turning angle distribution is uniform \citep[Eq. 13]{barnettmoorcroft2008}, where $z({\bf x},{\mathcal E})=\int_\Omega {\rm d}{\bf x}' \Phi({\bf x}'|{\bf x},\theta_0)w({\bf x}',{\mathcal E})$.  By numerically deriving the steady space-use distribution for our model, we show that this result breaks down when we include correlation in the movement process.  Figs. \ref{me_plots}a and \ref{me_plots}b compare the analytic result to the numerical one in the specific example of our Amazonian bird flock model, in the case $w(x,{\mathcal E})=w_b(x,\alpha,\beta)$ (see Eq. \ref{weight_fn_b}).  However, if we assume that the turning angle distribution is uniform, then the analytic solution is very similar (Figs. \ref{me_plots}b and \ref{me_plots}c).

% The Discussion should explain the significance of the results. Distinguish factual results from speculation and interpretation. 
\section*{Discussion}
%This should point out the significance of the results in relation to the reasons for doing the work, and place them in the context of other work.

We have constructed a step selection function (SSF) to test three hypotheses about the drivers behind Amazonian bird flock movement decisions.  We have shown that these flocks have a tendency to move towards areas covered by higher canopies, but move away from areas of higher ground.  The preference for higher canopies is likely to be due to the greater abundance of resources, through enhanced micro-climatic conditions in the understory and more foraging substrate \citep{bassetetal1992}.  Lower ground is likely to be preferred because it has a moister environment that can hold a higher insect biomass \citep{chanetal2008}.  

However, the flocks are just as likely to move back to a place that they have recently visited than one that they have not visited for a while.  This suggests that when they visit a tree, they do not deplete the resources as much as they can, but leave the tree in the knowledge that there is still plenty of food to be found there.  Whilst it may seem advantageous to stay at a tree as long as it is profitable to do so, in order to conserve energy \citep{houstonetal1993}, this frequent movement from tree to tree might be a tactic to avoid predators.  Alternatively, insects may temporarily be adopting cryptic behavior on the presence of birds, thus forcing the birds to move on quickly as insects become rapidly harder to find. 

We tested different functional forms for the selection weighting, something that is rarely done in literature on step selection functions but could be important \citep{leleetal2013}.  Although we would be surprised if the functional form were to change the outcome of hypothesis testing, it could very much affect the resulting parameters that are used to build the mechanistic model.  For example, an exponential effect of the canopy height vastly increases the relative attraction to very high canopies as compared with a power law effect, since this is effectively the difference between a linear and a logarithmic scaling (see the note after Eq. \ref{weight_fn_b}).  This has the potential to vastly change the predicted space use patterns.  Therefore it is vital to consider functional form when using step selection techniques to build mechanistic models.

Our SSF approach enabled us to run simulations that were used to predict the utilization distribution (UD) of a flock, thereby relating the small-scale movement decisions to the large-scale space use patterns.  While the resulting simulated UD captured certain qualitative aspects of the empirical data (Fig. \ref{sim_plots}), it overestimated the home range size.  In comparison, a straightforward random walk model, based on the empirical mean step length distribution, would give a normal distribution with the 90\% contour approximately 395m from the gathering point.  This contour would overlap the corresponding (outer) contour from Fig. \ref{sim_plots}, but would be circular, whereas the simulation contour is far from symmetric.  Therefore, though certain features of space use are being predicted by our model, there must be some other aspect of the birds' movement decisions keeping them far more spatially confined than our current model predicts.

We propose two plausible mechanisms that might explain this confinement.  First, these flocks are highly territorial \citep{develeystouffer2001}, so interactions with neighboring flocks may cause each flock to use less space than they would otherwise.  The mechanism of conspecific avoidance has been shown to give rise to  spatial confinement in various species of canid \citep{lewismurray1993,moorcroftetal2006,PHG3}.  These all deal with avoidance via scent marking, whereas territories in birds are defended via vocalizations and direct interactions \citep{munnterborgh1979}.  However, the generic modeling framework from \citet{PHG3} could be used to constructed coupled SSFs, whose weighting functions $w$ depend both upon the position of the individual and on interactions with neighbors.  These interactions may either be direct or mediated by vocal, visual or olfactory cues.  

Second, memory effects, with birds having a preference to move back towards places they have frequently visited, can cause spatial confinement.  Theoretical studies by \citet{briscoeetal2002} have described such a mechanism in wolf ({\it Canis lupus}) populations, and the general results of \citet{tanetal2001} show that memory can severely constrain the amount of area used in a given time period.  Though it is tricky to determine empirically what constitutes a bird's cognitive map of the environment, it is generally considered that memory is an important factor in the spatial confinement and site-fidelity of many animals \citep{smouseetal2010}.   

% Master equation; Moorcroft and Barnett
By turning our SSF into a master equation for the spatio-temporal probability distribution of the flock's position, we compared our results to a recent approximate analytic prediction by \citet{barnettmoorcroft2008} that applies when the turning angle distribution is uniform.  However, their results fail whenever there is correlation in the animal's movement at any time-scale, a fact noted in \citet{barnettmoorcroft2008} but may not be clear to those only familiar with the more ecologically-motivated paper of \citet{moorcroftbarnett2008}.  The more the correlation, the worse the prediction is likely to be, so it is necessary to take care when applying these results to empirical data.  Though the correlation in the birds' movement greatly affected the movement pattens, when we removed any intrinsic correlation from our movement model, the predictions of \citet{barnettmoorcroft2008} were visually very good (Fig. \ref{me_plots}).

%Intrinsic correlation in movement processes can cause a pattern similar to that which would arise from localized advection, if the movement is also affected by environmental covariates, which may explain the disparity between the results of \citet{barnettmoorcroft2008} and ours.  For example, if there are two patches of good resources at positions ${\bf x}_a$ and ${\bf x}_b$, which a flock may have a tendency to move between, then if the flock also has intrinsic correlation in its movement then it would be more likely to be found at a position just to the other side of ${\bf x}_b$ from ${\bf x}_a$, even though this position may be resource poor, than at some position just as close to ${\bf x}_b$, and of just as good quality, but in an orthogonal direction to ${\bf x}_b-{\bf x}_a$.  This causes the quality of resource patches on space use to be `smeared out' in certain directions, dependent on the geometry of the resource distribution.  When we remove any intrinsic correlation from our movement model, this smearing-out effect is no longer there.   Therefore the predictions of \citet{barnettmoorcroft2008} are very good (Fig. \ref{me_plots}).

The application of these models could provide basis for informed management decisions for a subset of the avian community that is known to be very sensitive to forest disturbances. By providing information on how a combination of two important habitat features influences habitat use and how these flocks anchor their home ranges, this would allow for more realistic estimations of areas that are more important to these species. Also, the drivers related to resource abundance and renewal provide important insights on the nature of the relationship of cursorial insectivorous birds and their resource, a topic that has challenged researchers for years \citep{sherry1984, sekercioglusetal2002}.  

%While we only focus on three specific types of environmental drivers, there are myriad factors, that can affect the space use of animals, depending on the species and habitat.  As well as the aspect of foraging needs, territorial interactions have been shown to give rise to spatial confinement in coyotes ({\it Canis latrans}) \citep{moorcroftetal2006} and red foxes ({\it Vulpes vulpes}) \citep{PHG3}.  Memory effects, while tricky to measure empirically, have also been shown theoretically to cause a certain amount of confinement \citep{tanetal2001, briscoeetal2002}.  Anthropogenic linear features can also affect movement in large mammals \citep{fortinetal2005, mckenzieetal2012}, so are likely to have an effect on space use.  Similarly predator-prey interactions are well-known to affect spatial population structure via animal movement processes \citep{lewismurray1993, latombeetal2013}.  Therefore we are confident that our techniques can be readily generalized to many species in a wide variety of environments.

\section*{Acknowledgements}
%In addition to acknowledging collaborators and research assistants, include relevant permit numbers (including institutional animal use permits), acknowledgment of funding sources, and give recognition to nature reserves or other organizations that made this work possible.
This study was partly funded by NSERC Discovery and Acceleration grants (MAL, JRP).  MAL also gratefully acknowledges a Canada Research Chair and a Killam Research Fellowship.  JRP also acknowledges a Worldwide Universities Network Research Mobility Award.  KM would like to acknowledge the Biological Dynamics of Forest Fragments Project (BDFFP) staff for providing logistic support; J. Lopes, E.L. Retroz, P. Hendrigo, A. C. Vilela, A. Nunes, B. Souza, M. Campos for field assistance; M. Cohn-Haft for valuable discussions. Funding for the research was provided by US National Science Foundation grant LTREB 0545491 to PCS and by the AOU 2010 research award to KM. This article represents publication no. xxx in the BDFFP Technical Series. This is contribution no. xx in the Amazonian Ornithology Technical Series of the INPA Zoological Collections Program. This manuscript was approved for publication by the Director of the Louisiana Agricultural Experiment Station as manuscript 2014-xxx-xxxx. LIDAR images for canopy height models and digital elevation models were provided by Scott Saleska (University of Arizona) and Michael Lefsky (Colorado State University).  We are grateful to Greg Breed, the Lewis Lab, and various other colleagues for helpful discussions and comments.

\section*{Data accessibility}
%Please state where you have deposited the raw data underlying your analyses. It will need to include the name of the repository (e.g. Dryad, figshare, GenBank etc.) and location of the data (i.e DOI). For authors archiving at Dryad, we can facilitate the process when your paper is accepted. \\
Data can be found in Dryad doi:10.5061/dryad.47jh1

\bibliographystyle{jae}%Compile with jae.bst style file

%\bibliography{refs}% your .bib file(s)
\newpage
\section*{Table captions}
\noindent \textbf{Table~1.} Results of hypothesis testing.  The first column is number of the test, as given in the introduction.  This test finds the best fit parameter given in the second column. The third column denotes the weighting function used for the test (see Eqs. \ref{weight_fn_a}, \ref{weight_fn_b}, \ref{weight_fn_c} and \ref{weight_fn_d}) and the fourth gives the value of the time $\tau$ between successive position measurements in the data. The fifth column shows the value of the parameter that fits the data best ($\pm$standard error), with a $p$-value from the likelihood ratio test (see Methods) given in the sixth column and the results of a 1\% significance test in the final column (note that a 5\% test would give identical results).
\newpage
%\begin{table}[h!]
%  \caption{}
%  \label{TabENFA}
%  \begin{center}
%    \begin{tabular}{lrrr}
%      \hline
%      Condition & Treatment 1 & Treatment 2 & Treatment 3 \\
%      \hline
%      A & value & value & value \\
%      B & value & value & value \\
%      C & value & value & value \\
%      D & value & value & value \\
%      E & value & value & value \\
%      \hline
%    \end{tabular}
%  \end{center}
%\end{table}
\begin{table}[!ht]
\begin{flushleft}
\caption{\small Results of hypothesis testing.  The first column is number of the test, as given in the introduction.  This test finds the best fit parameter given in the second column. The third column denotes the weighting function used for the test (see Eqs. \ref{weight_fn_a}, \ref{weight_fn_b}, \ref{weight_fn_c} and \ref{weight_fn_d}) and the fourth gives the value of the time $\tau$ between successive position measurements in the data. The fifth column shows the value of the parameter that fits the data best ($\pm$standard error), with a $p$-value from the likelihood ratio test (see Methods) given in the sixth column and the results of a 1\% significance test in the final column (note that a 5\% test would give identical results).}  
\label{hypothesis_test_results}
\end{flushleft}
\begin{tabular}{  l  l  l  l  l  l  }				
\hline			
  Test & Parameter & $w$-function & $\tau$ (mins) & Best fit & $p$-value  \\
\hline  
%  1 & $\alpha$ & $w_a$ & 1 & 0.00665 & 0.0031 & Yes \\
%  2 & $\beta$ & $w_a$ & 1 & 0.0247 & 0.000000027 & Yes \\
%  3 & $G$ & $w_c$ & 1 & 1.00 & 1.0 & No  \\
   &  &  &  &  &  \\
  1 & $\alpha$ & $w_b$ & 1 & 0.095$\pm$0.037 & 0.0038 \\
%  2 & $\beta$ & $w_b$ & 1 & 1.658 & 0.000000020 & Yes\\
   &  &  &  &  &  \\
  2 & $\beta$ & $w_b$ & 1 & 1.658$\pm$0.345 & $< 0.001$\\
   &  &  &  &  &  \\
  3 & $G$ & $w_d$ & 1 & 1.00 & N/A \\
%  1 & $\alpha$ & $w_a$ & 5 & 0.0112 & 0.0053 & Yes \\
%  2 & $\beta$ & $w_a$ & 5 & 0.0242 & 0.00011 & Yes\\
%  3 & $G$ & $w_c$ & 5 & 1.00 & 1.0 & No \\
%  1 & $\alpha$ & $w_b$ & 5 & 0.227 & 0.00014 & Yes \\
   &  &  &  &  &  \\
  1 & $\alpha$ & $w_b$ & 5 & 0.227$\pm$0.065 & $< 0.001$  \\
%  2 & $\beta$ & $w_b$ & 5 & 1.697 & 0.000066 & Yes\\
   &  &  &  &  &  \\
  2 & $\beta$ & $w_b$ & 5 & 1.697$\pm$0.436 & $< 0.001$ \\
   &  &  &  &  &  \\
  3 & $G$ & $w_d$ & 5 & 1.00 & N/A  \\
\hline 
\end{tabular}
\end{table}

\newpage
\section*{Figure captions}
\begin{figure}[h!]
  \caption{}
  \label{slta}
  \begin{center}
    \includegraphics[width=120mm]{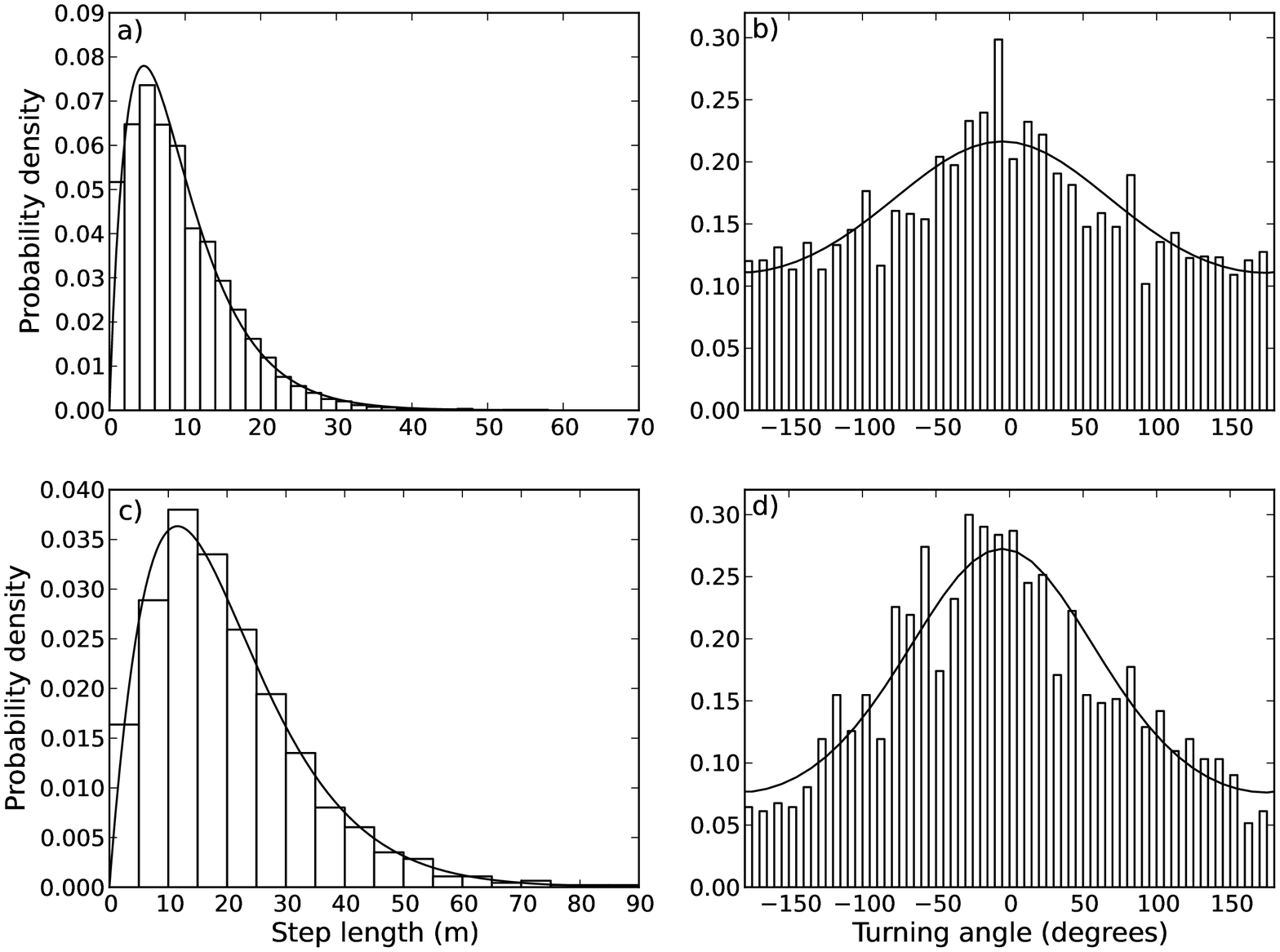}
  \end{center}
\end{figure}

\begin{figure}[h!]
  \caption{}
  \label{sim_plots}
  \begin{center}
    \includegraphics[width=120mm]{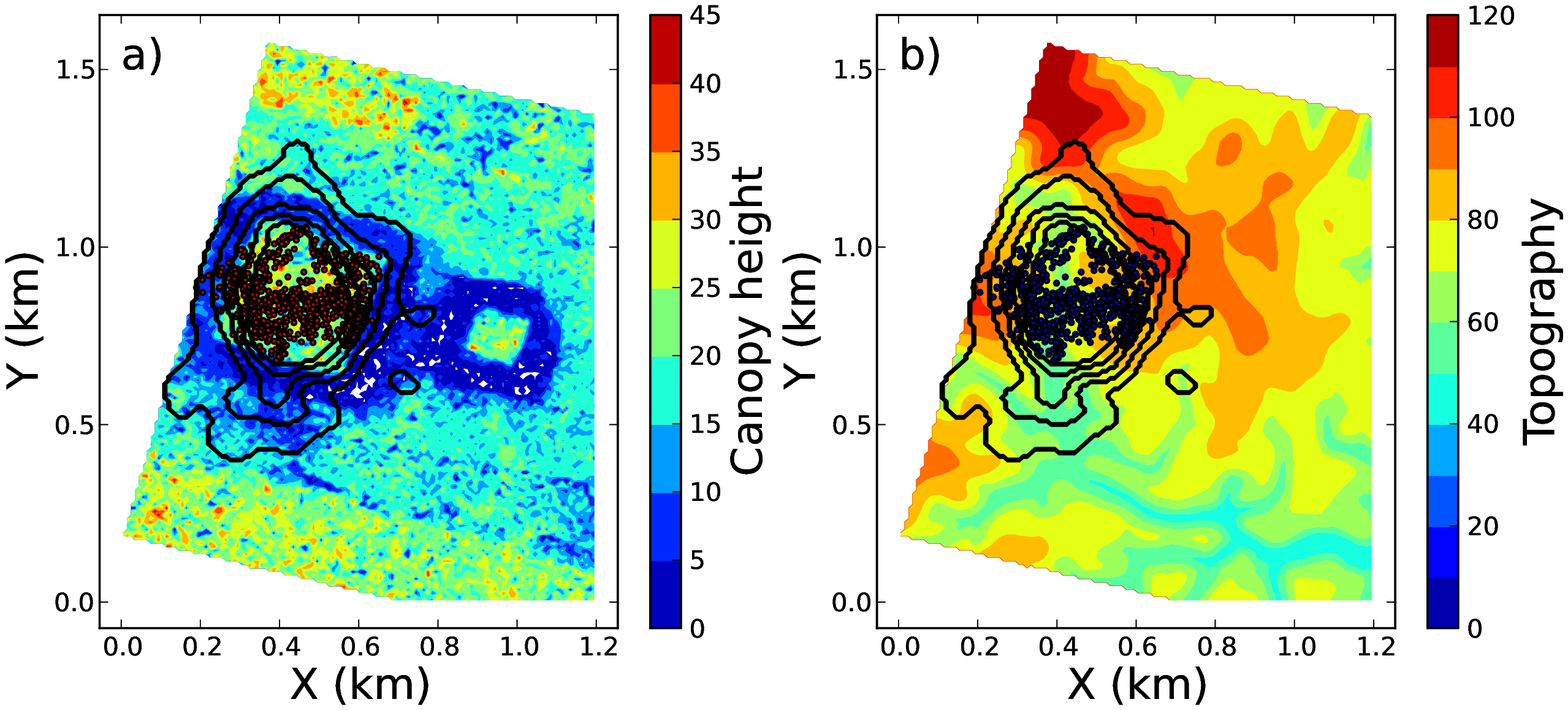}
  \end{center}
\end{figure}

\begin{figure}[h!]
  \caption{}
  \label{me_plots}
  \begin{center}
    \includegraphics[width=120mm]{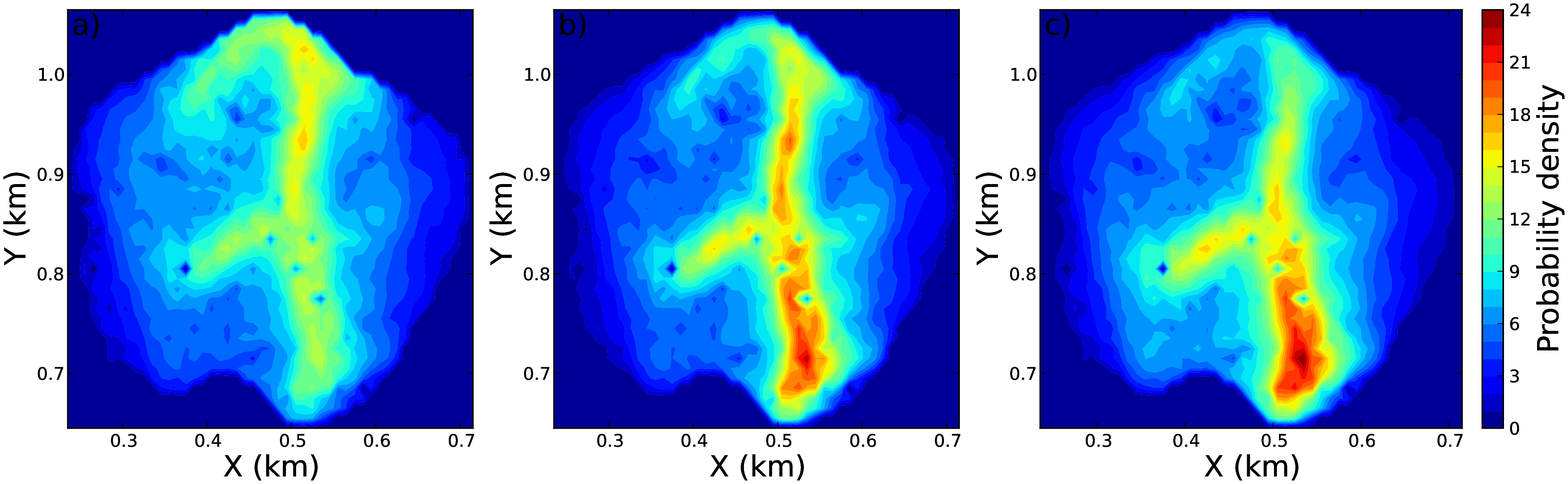}
  \end{center}
\end{figure}

\noindent \textbf{Figure~1.} {\bf Step length and turning angle distributions.}  Panel (a) shows the empirical step length distribution (bars) for data where the temporal resolution is $\tau=1$ minute, together with the best fit Exponentiated Weibull distribution (solid curve).  The latter is given in Eq. \ref{step_length_exp_weibull}, with $a=1.06$, $b=6.90$ and $c=1.82$.  The bars in panel (b) denote the empirical turning angle distribution for the same data, whereas the curve denotes the best fit von Mises distribution, given in Eq. \ref{turning_angles} with $k=0.336$.  Panels (c) and (d) are analogous to (a) and (b) respectively, except they use the data set where $\tau=5$ minutes, rather than $\tau=1$ minute.  Here, $a=1.26$, $b=17.2$, $c=1.55$ and $k=0.637$. \\
\noindent \textbf{Figure~2.} {\bf Plots of simulated and real data.} Both panels shows the empirical data for one flock (dots) together with the 50\%, 60\%, 70\%, 80\% and 90\% kernel density estimation curves for the simulated data (black curves).  See the Methods section for details on how the simulations were performed.  The colors underlying panel (a) denote the canopy height, whereas in panel (b) they give the topography, i.e. height of the ground above sea level. \\
\noindent \textbf{Figure~3.} {\bf Exact and approximate steady state solutions of the master equation.} Panel (a) shows the numerical steady state solution of our master equation (Eq. \ref{ME}) with $w=w_b$ (Eq. \ref{weight_fn_b}) and the parameters that best fit the data (see Fig. \ref{slta} and Table \ref{hypothesis_test_results}).  The numbers on the axes correspond to those in Fig. \ref{sim_plots} for ease of comparison.  The analytic solution, given in \citet[Eq. 13]{barnettmoorcroft2008}, is given in panel (b).  Though there are some similarities between panels (a) and (b), the approximation is evidently not particularly good.  However, when we replace the von-Mises turning angle distribution with a uniform distribution, the numerical steady state solution of Eq. \ref{ME} (panel c) is visually very close to that of panel (b), as expected.
\newpage
\newpage
\newpage
\end{document}